\documentclass[twocolumn]{aastex63}

\submitjournal{ApJL}

\shorttitle{Signatures of AGN induced metal loss in the stellar population}
\shortauthors{Camps-Fariña et al.}
\graphicspath{{./}{figures/}}

\begin{document}

\title{Signatures of AGN induced metal loss in the stellar population}

\correspondingauthor{A. Camps-Fariña}
\email{acamps@astro.unam.mx}

\author{A. Camps-Fariña}
\affiliation{Instituto de Astronom\'ia\\ Universidad Nacional Aut\'onoma de M\'exico\\ Apartado Postal 70-264\\ CP 04510 Ciudad de M\'exico, M\'exico\\}

\author{S. F. Sánchez}
\affiliation{Instituto de Astronom\'ia\\ Universidad Nacional Aut\'onoma de M\'exico\\ Apartado Postal 70-264\\ CP 04510 Ciudad de M\'exico, M\'exico\\}

\author{L. Carigi}
\affiliation{Instituto de Astronom\'ia\\ Universidad Nacional Aut\'onoma de M\'exico\\ Apartado Postal 70-264\\ CP 04510 Ciudad de M\'exico, M\'exico\\}

\author{E. A. D. Lacerda}
\affiliation{Instituto de Astronom\'ia\\ Universidad Nacional Aut\'onoma de M\'exico\\ Apartado Postal 70-264\\ CP 04510 Ciudad de M\'exico, M\'exico\\}

\author{R. García-Benito}
\affiliation{Instituto de astrof\'isica de Andaluc\'ia (CSIC) PO Box 3004, 18080 Granada, Spain\\}

\author{D. Mast}
\affiliation{Observatorio Astron\'{o}mico de C\'{o}rdoba, Laprida 854, X5000BGR, C\'{o}rdoba, Argentina.\\}
\affiliation{Consejo de Investigaciones Cient\'{i}ficas y T\'ecnicas de la Rep\'ublica Argentina, Avda. Rivadavia 1917, C1033AAJ, CABA, Argentina.\\}

\author{L. Galbany}
\affiliation{Institute of Space Sciences (ICE, CSIC), Campus UAB, Carrer de Can Magrans, s/n, E-08193 Barcelona, Spain.\\}

\author{J. K. Barrera-Ballesteros}
\affiliation{Instituto de Astronom\'ia\\ Universidad Nacional Aut\'onoma de M\'exico\\ Apartado Postal 70-264\\ CP 04510 Ciudad de M\'exico, M\'exico\\}

\begin{abstract}
One way the AGN are expected to influence the evolution of their host galaxies is by removing metal content via outflows. In this article we present results that show that AGN can have an effect on the chemical enrichment of their host galaxies using the fossil record technique on CALIFA galaxies. We classified the chemical enrichment histories of all galaxies in our sample regarding whether they show a drop in the value of their metallicity. We find that galaxies currently hosting an AGN are more likely to show this drop in their metal content compared to the quiescent sample. Once we separate the sample by their star-forming status we find that star-forming galaxies are less likely to have a drop in metallicity but have deeper decreases when these appear. This behavior could be evidence for the influence of either pristine gas inflows or galactic outflows triggered by starbursts, both of which can produce a drop in metallicity.
\end{abstract}




\section{Introduction} \label{sec:intro}

Feedback from supermassive black holes (SMBH) in active galaxies plays a key role in the evolution of galaxies. The radiation and kinetic energy they inject into the surrounding interstellar medium (ISM) can shut off star formation \citep[][]{Silk1998}. Active galactic nuclei (AGN) also have an effect on the dynamics of the host galaxy, especially in the central regions with a well known correlation between the SMBH mass and the dispersion of the stellar velocities in the bulge \citep{Ferrarese2002, Tremaine2002, Kormendy2013}.

The injection of kinetic energy in the central region of the host galaxy is capable of producing massive outflows which strip it of material. The gas in the center of galaxies tends to be richer than at further galactocentric distances (for stellar masses above 10$^{10}$ M$_\odot$), so a massive nuclear outflow can strip the galaxy of a large quantity of metals \citep{Nesvadba2006, Simionescu2008, Kirkpatrick2009, Rodriguez-Gonzalez2011, Cicone2014, Robles-Valdez2017}

Outflows have been shown to be an important process in the context of galaxy evolution, in particular for the evolution of the metal content of galaxies, and as such are supposed to be a major factor in shaping the mass-metallicity relation \citep[MZR, ][]{Tremonti2004, Brooks2007, Kobayashi2007, deRossi2007, Dave2011} as well as in the enriching of the intergalactic medium \citep[e.g.,][]{Oppenheimer2008, Oppenheimer2009}

\citet{Torrey2014} study how the feedback from, among others, AGN can affect the stellar content of their host galaxies by performing cosmological simulations and comparing the resulting galaxies to observational data at several redshifts. \citet{Calura2009} perform a similar study using a semi-analytical model which applies known physics and relations including those of AGN to predict how the abundance of several elements evolves over cosmic time. An observational analysis of the impact AGN have on the chemical evolution of galaxies such as the work we present here can be an important tool in furthering these kinds of studies.

In this article we present an analysis of the chemical enrichment histories (ChEH) of the AGN sample within the CALIFA dataset and compare it to those of the quiescent sample. We also perform the same analysis but comparing the galaxies depending on their star formation. These comparisons allow us to check whether the presence of an AGN or a massive star formation episode can influence the metallicity of the stellar populations in the host galaxy. The article is structured as follows: In Section \ref{sec:sample} we present the data we used and how we produced the samples used for the analysis, which is described in Section \ref{sec:analysis}. Finally, in Section \ref{sec:results} we present the results.

\section{Sample and Data} \label{sec:sample}
We use the Calar Alto Legacy Integral Field Area Survey (CALIFA) as the base for the study. The base sample consists of all galaxies included in the DR3 of the CALIFA survey \citep{Walcher2014} as well as additional galaxies from extended surveys \citep{Sanchez2016a} such as the PMAS/Ppak Integral-field Supernova hosts COmpilation \citep[PISCO; ][]{Galbany2018}.

The CALIFA survey uses the PMAS/PPAK integral field unit (IFU) spectrophotometer, which for the data used in this study was configured in the V500 setup. In this setup it covers a wavelength range from 3745\AA{} to 7200 \AA{} with a spectral resolution of R$\sim850$ corresponding to about FWHM$\sim$6.5\AA{} over the spectral range.

This data is reduced using the pyR3D \citep[][]{Sanchez2016a} software which is an improved version of the previous reduction pipeline \citep[][]{Sanchez2006} which uses standard procedures to extract and reduce the spectra and construct a data cube consisting of a spectrum at each spatial position. The field of view (FoV) of the cubes is 74" X 64" and the galaxies comprising the sample are selected such that the FoV covers at least 2.5 times the effective radius (Re) in all the galaxies. The spatial resolution of the data is 2.5$\arcsec$/FWHM, resulting from applying a dithering technique to the observations which also mitigates the effects of the gaps between the fibers.

This base sample is then refined to include only those observations which have good spectroscopic data as well as galaxies which have an inclination below 70º, to avoid the uncertainties in the stellar population fitting results that arise in edge-on galaxies.
The final sample consists of 668 galaxies and is divided into the quiescent and AGN host subsamples. The detection of the AGN follows the analysis performed by \cite{Lacerda2020} which uses the BPT diagram and equivalent width in H$\alpha$ (EW$_{ \mathrm{H}\alpha }$) to classify the galaxies. In this work we accept galaxies which fulfill the criteria for EW$_{ \mathrm{H}\alpha }$ and two of the three BPT criteria ([OIII]/H$\beta$ versus either [NII]/H$\alpha$, [SII]/H$\alpha$, or [OI]/H$\alpha$) yielding 49 AGN hosts and 619 quiescent galaxies.

 \begin{figure}
 	\includegraphics[width=\linewidth]{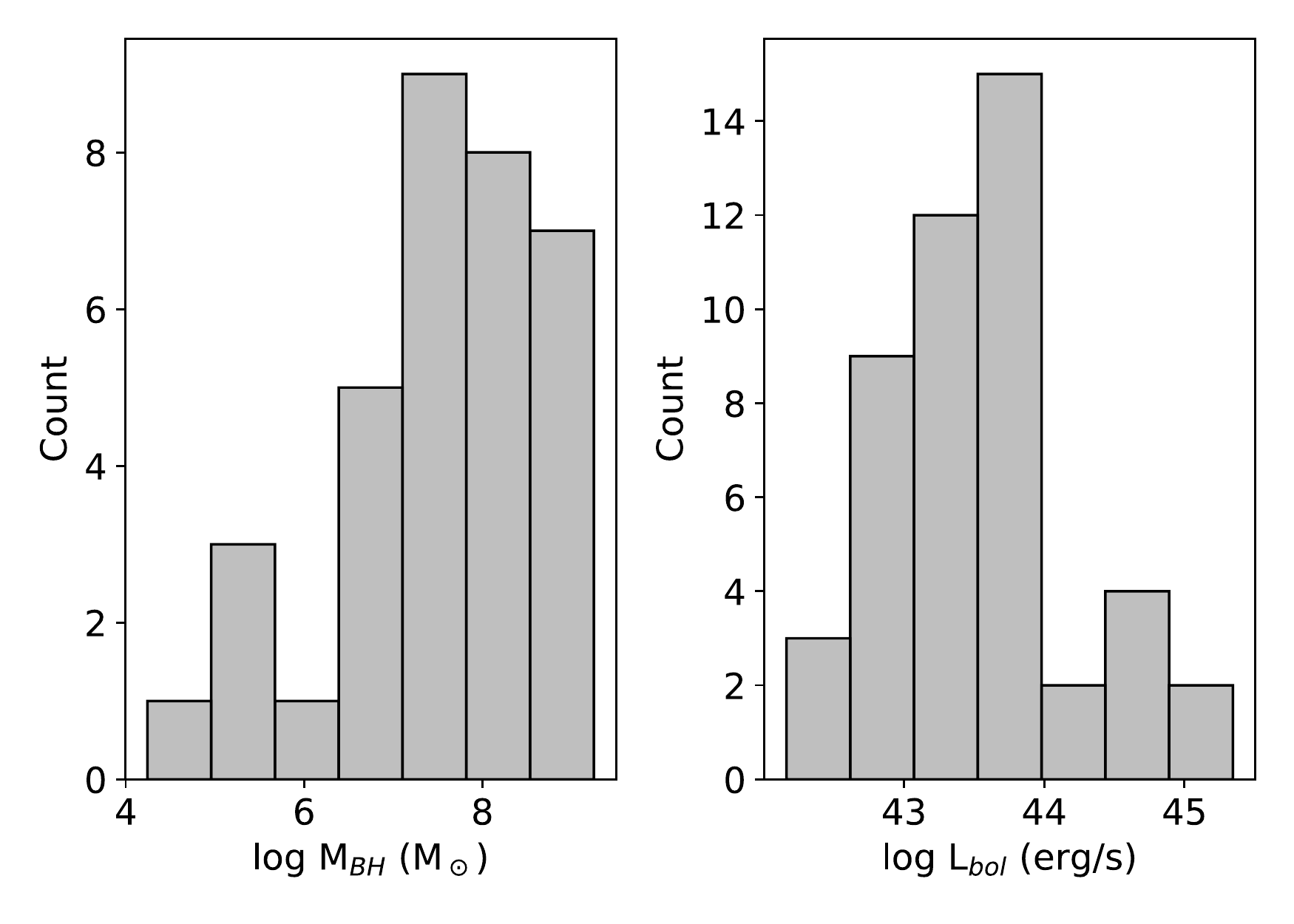}
     \caption{Distribution of the AGN in our sample in terms of their estimated black hole mass (M$_{BH}$, left) and luminosity (L$_{BH}$, right). The M$_{BH}$ was calculated using the relations from \cite{Greene2020} and the L$_{BH}$ was derived using the calibration using [OIII] and [OI] from \cite{Netzer2009}.}
     \label{fig:agn_mbh}
 \end{figure}

In Figure \ref{fig:agn_mbh} we show the distribution of black hole masses and luminosities, calculated using the relations from \cite{Greene2020} and \cite{Netzer2009} respectively. The distributions show that the SMBH in our sample are moderate in terms of their mass and luminosity. It is important to mention that our selection is not complete in terms of the AGN population as it is detected only using optical methods. As such we will miss many obscured or weaker AGN \citep[e.g.][]{Azadi2017, Koss2017}.

\section{Analysis} \label{sec:analysis}
In order to obtain information on the underlying stellar population such as the stellar mass, metallicity and star-forming history FH) which we use in this study we apply the fossil record technique to the IFU data. By fitting a set of stellar population templates to an observed spectrum with the emission lines subtracted we can recover the weights of the stellar populations.
We use the analysis pipeline \textsc{pipe3d}  \citep{Sanchez2016a,Sanchez2016b} with the stellar population library GSD156 \citep{CidFernandes2013}. The GSD library is composed using the observational MILES stellar population spectra \citep{Vazdekis2010} for populations older than 63 Myr and models by \cite{GonzalezDelgado2005} for younger stars. It consists of 156 templates covering four metallicity values and 39 ages distributed in a pseudo-logarithmic sampling such that the interval between templates in age is larger for older populations.

Using the weights for the stellar populations provided by \textsc{pipe3d} we can obtain the ChEH for each galaxy by averaging the metallicity of the populations that contribute light at each look-back time (LBT). We can also measure the SFH as the change in stellar mass over time. The mass fraction of each population is corrected to account for the loss of stars that occurs as the population ages \citep[see][]{Sanchez2016b}. Since metallicity is an intensive parameter we need to weigh the average by the contribution of each population, which can be done using the luminosity or the stellar mass. The metallicity values shown here are weighted by the luminosity of the populations which biases the results somewhat to the younger populations. As AGNs are relatively short-lived phenomena (in terms of galaxy evolution) weighing by luminosity is advantageous in order to observe the effects that they have on the evolution of the host galaxy. 

The resulting ChEHs are separated into bins depending on their mass and activity status and then averaged as to obtain a representative ChEH of each bin. This is done following the procedure detailed in \citet{Camps-Farina2021} to avoid issues from the different redshift values of the galaxies.

\section{Results} \label{sec:results}

 \begin{figure*}
 	\includegraphics[width=\linewidth]{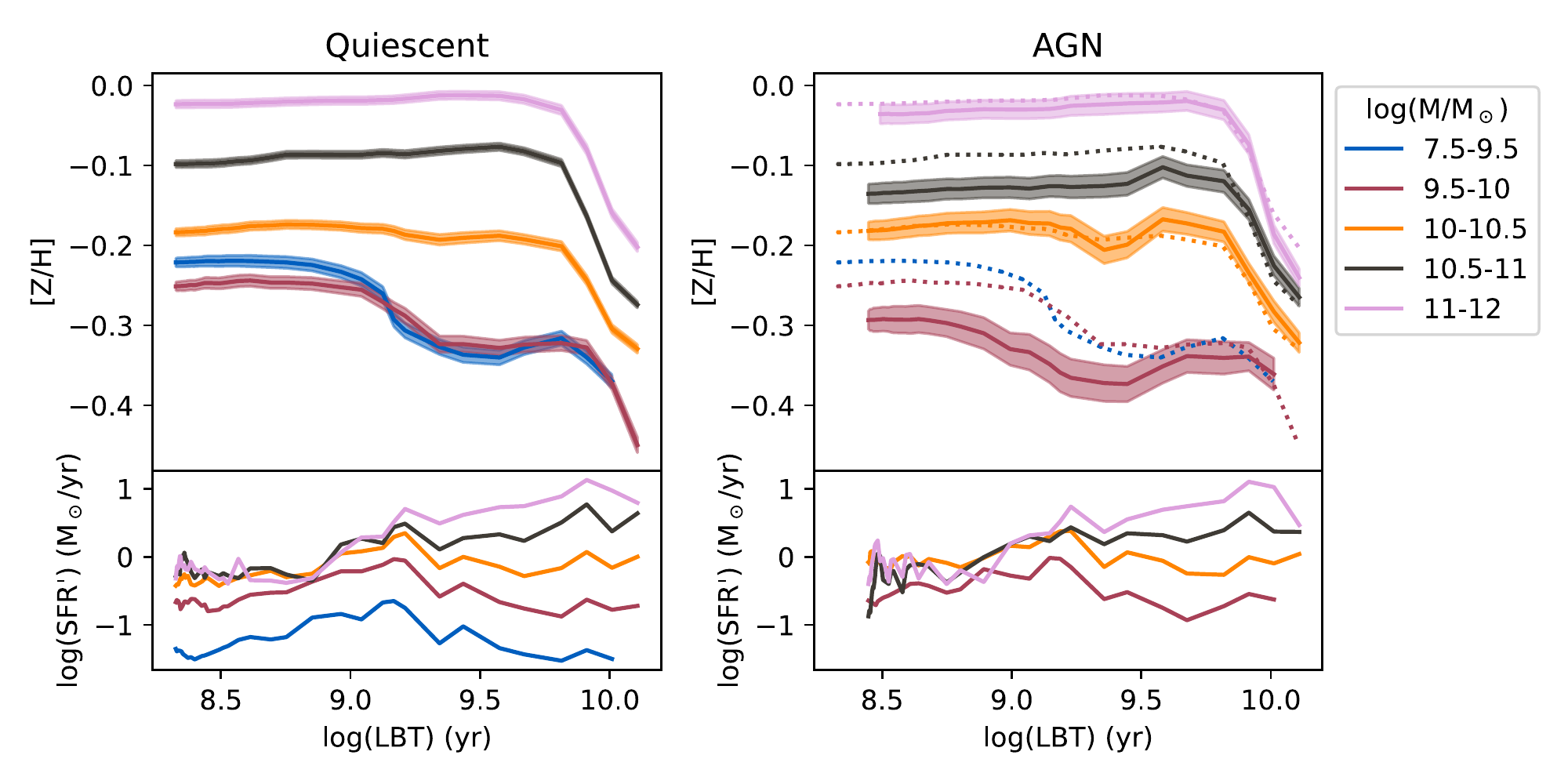}
     \caption{On the left, the ChEH for the quiescent portion of the CALIFA sample (top panel) with the SFH on the bottom panel. On the right the same but for the galaxies currently hosting an AGN with the quiescent ChEHs overlaid as dotted lines. The colors correspond to mass bins in stellar mass of the galaxies. The SFH is computed as the derivative of the mass assembly function \citep[see][]{Camps-Farina2021}. The shaded areas correspond to the error of the mean.}
     \label{fig:zh_agn}
 \end{figure*}
 
In Figure \ref{fig:zh_agn} we show a comparison between the averaged ChEHs and SFHs of the quiescent and AGN sub-samples of the CALIFA survey. The ChEHs are averaged within mass bins because AGN host galaxies are more massive on average compared to the quiescent sample. If we do not separate into mass bins it could introduce a bias in the chemical evolution that is not due to the presence of an AGN. The usage of mass bins is also in accordance with \citet{Camps-Farina2021}.

The ChEH for the lowest mass bin present in the AGN host sample ($10^{9.5-10}$ M$_\sun$) shows a lower current metallicity compared to the quiescent sample. The $10^{10-10.5}$ M$_\sun$ has similar values but shows a dip in at about $10^{9.4}$ yr in LBT. At a similar LBT ($\sim 10^{9.5}$ yr) the $10^{10.5-11}$ M$_\sun$ mass bin shows a decrease in metallicity not present for the quiescent sample. The most massive bin shows very little difference. The SFHs show no obvious differences between the two groups.

The decrease shown for the $10^{10.5-11}$ M$_\sun$ mass bin is consistent with dilution of the metal content by a metal rich outflow which could be powered by an AGN. It is also interesting to note that the ChEH at the earlier LBT (up to $\sim 10^{9.8}$ yr) has the same values and shape for either quiescent and AGN host galaxies for this mass bin before the metallicity drops. This is also observed for the most massive bin, but the difference in metallicity is very low and within the error bar. In general, the effects of the presence of an AGN are small ($\lesssim 1$ dex) and not completely consistent among the mass bins. An explanation for this is that metallicity decreases induced by outflows would not necessarily occur at the same cosmic time, which would dilute the signature.

It is in the individual ChEHs, then, that we may discern whether the differences we have observed are spurious or not. In order to assess this we classified all galaxies in the full sample (AGN and quiescent) in terms of the shape of their ChEH. Specifically, we selected those galaxies that showed a clear decrease in their metallicity after the initial enrichment.

 \begin{figure}
 	\includegraphics[width=\linewidth]{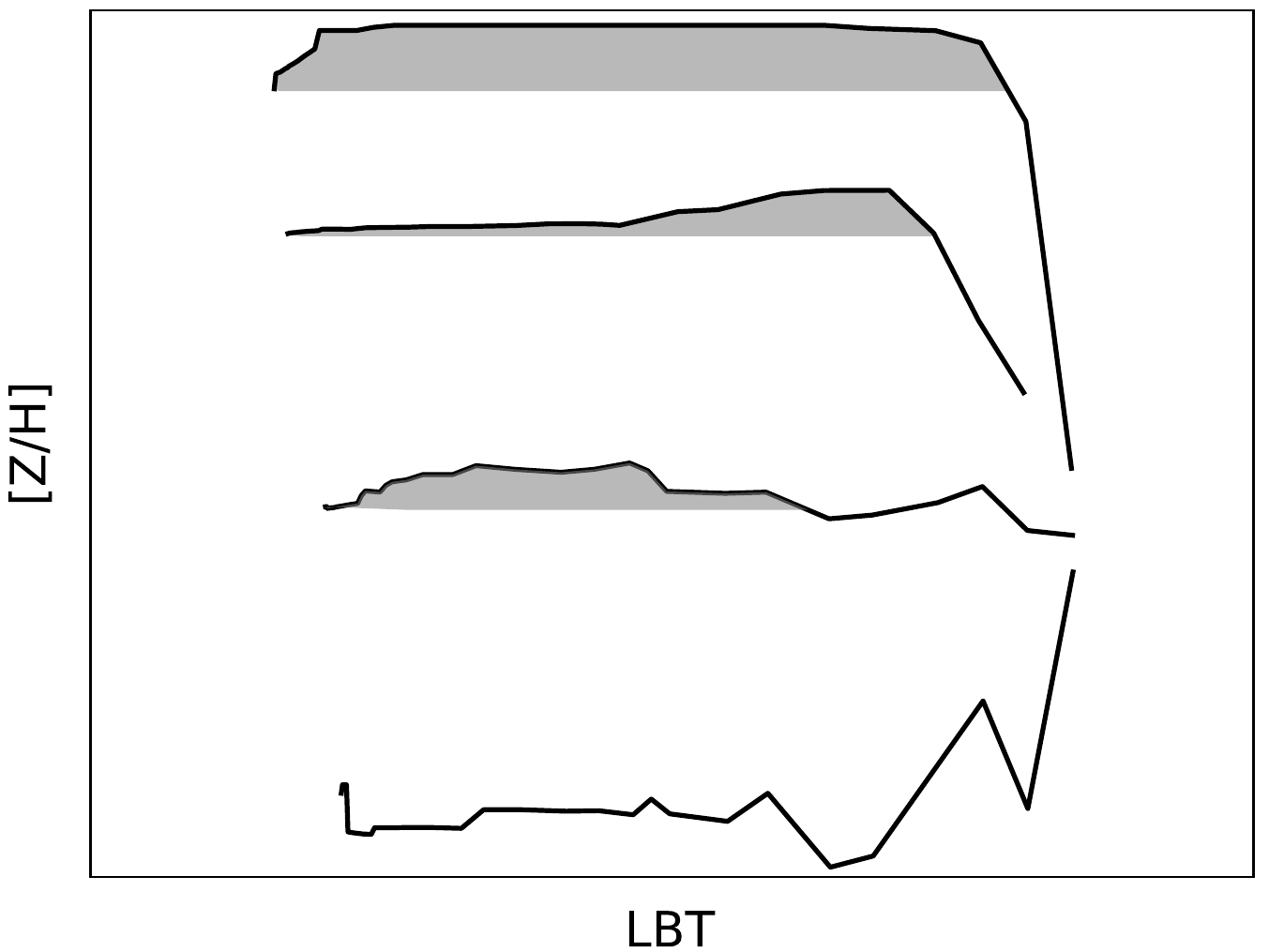}
     \caption{A few examples illustrating the criteria for classifying the shapes of the ChEHs in the sample, with the shaded areas showing the decrease in metallicity after its maximum value. The top two ChEHs are considered to show a clear decrease in metallicity after the initial enrichment. The bottom two are examples of ChEHs which fulfill a numerical criteria of a decrease in the value of the metallicity but which are not considered in this work. In the case of the third ChEH the decrease is similar to the fluctuations in metallicity over the galaxy's lifetime which might not be physical. The bottom ChEH is a more extreme case of the same criterion with larger fluctuations.}
     \label{fig:examples}
 \end{figure}

In Figure \ref{fig:examples} we show a few examples to illustrate the criteria employed in the classification. In general terms we select galaxies whose ChEH shows a smooth evolution except for a clear decrease in the value of the metallicity which cannot be attributed to the intrinsic fluctuations of the metallicity values. These criteria are easy for the human eye to discern but are not trivially implemented in numerical code. Since the number of galaxies in the sample is manageable for a manual examination we classified the ChEHs by eye.
Naturally, to prevent any biases in the determination all galaxies were classified without information as to whether they host an AGN.

\begin{deluxetable*}{lccccc}
\tablenum{1}
\tablecaption{Distribution and properties of galaxies with a decrease}
\tablewidth{0pt}
\tablehead{
\colhead{Subgroup} & \colhead{Rate of $\Delta$[Z/H]}  & \colhead{Mean $\Delta$[Z/H]} &
\colhead{Median $\Delta$[Z/H]} & \colhead{Mean $\Delta$LBT} & \colhead{Median $\Delta$LBT} \\
 \colhead{} & \colhead{\%}  & \colhead{dex} &
 \colhead{dex} & \colhead{Gyr} & \colhead{Gyr}
}
\decimalcolnumbers
\startdata
AGN & 71 $\pm$ 10  & 0.059 $\pm$ 0.010  & 0.037 $\pm$ 0.010 & 3.6 $\pm$ 0.3 & 3.55 $\pm$ 0.3 \\
Quiescent & 44.6 $\pm$ 0.5  & 0.051 $\pm$ 0.002 & 0.042 $\pm$ 0.002 & 3.3 $\pm$ 0.1 & 3.55 $\pm$ 0.1 \\
& & & & & \\
RG & 56 $\pm$ 3 & 0.030 $\pm$ 0.002 & 0.022 $\pm$ 0.002 & 3.2 $\pm$ 0.1 & 3.55 $\pm$ 0.1 \\
GVG & 58 $\pm$ 5 & 0.059 $\pm$ 0.004 & 0.054 $\pm$ 0.004 & 3.2 $\pm$ 0.2 & 3.55 $\pm$ 0.2 \\
SFG & 33  $\pm$ 2 & 0.084 $\pm$ 0.005 & 0.075 $\pm$ 0.005 & 3.6 $\pm$ 0.2 & 3.55 $\pm$ 0.2 \\
\enddata
\tablecomments{Relevant properties for galaxies which show a decrease in [Z/H] in their ChEH. We separate the sample into groups depending on (i) nuclear activity (AGN and Quiescent) and (ii) star forming status (RG, GVG, SFG). We show the percentage of galaxies that show a decrease in metallicity (2), the mean (3) and median (4) values of the decrease in [Z/H] and the mean (5) and median (6) values of the time interval between the LBT at which the galaxy had its maximum value of [Z/H] and the LBT at which its light was emitted (equivalent to the redshift). For the (3) to (6) properties they were calculated only within the galaxies that both belong to a subgroup and show a decrease in metallicity.}
\label{tab:props}
\end{deluxetable*}

In Table \ref{tab:props} we show statistical properties related to the incidence of a decrease in metallicity in the ChEHs. It is in the incidence of a decrease in metallicity that the difference between AGN hosts and quiescent galaxies becomes clear with a much higher incidence of metal content decrease in AGN hosts at 71\% compared to 45\% for quiescent galaxies. The error for the percentages was calculated using the bootstrapping technique, by randomly selecting a new sample 100 times from the original one and recalculating the percentages. Since AGN hosts are much less numerous than quiescent galaxies they show a higher statistical error.

The fraction of AGN hosts with decreasing metallicity is statistically much larger than this fraction in any of the explored subgroups. Fisher's exact test between the quiescent and AGN samples yields a p-value of 0.0003 for the difference in incidence being a random result. This suggests that AGNs remove metals from their host galaxies to the extent that it influences their chemical history. There are several caveats to this interpretation however, first of which is that we are observing a statistical difference and not a direct correlation. What we call metal loss is the result of the competition between processes that dilute the gas which forms stars such as metal-rich outflows and processes that enrich the gas, mainly by-products of star-formation. Even a galaxy which is losing metals via metal-rich outflows would not show the feature we are considering as long as the enrichment outpaces the dilution. This fits with the results in terms of how galaxies with on-going star-formation have a lower incidence of the metallicity drop.

Finally, we need to consider the time-scales involved. We are measuring the metal content of stellar populations formed in the past and comparing it to the current activity of the nucleus. It is very unlikely that current AGN have had time to influence the stellar population to the extent that it influences the global metallicity values, even those weighted by luminosity, and therefore the metallicity decreases should not be directly related to the current AGN. It is for this reason that it is important to stress that these results need to be interpreted in a statistical sense: Our results can be explained as long as galaxies which currently host an AGN are more likely to have had an episode in their recent history (a few Gyr) in which a significant amount of metals was removed.

\citet{Stasinska2015} shows that for massive galaxies between $10^{10}$ and $10^{12}$ M$_\sun$ the AGN is intermittently active for 1-5 Gyr. This estimation agrees with theoretical models motivated by the physics of the accretion \citep{Martini2004}. \citet{Hopkins2009} use Eddington ratios to constrain the life-times of quasars. For high Eddington ratios ($0.1 < \lambda_{Edd} < 1$) they estimate lifetimes of $\sim10^8$ yr but for low values ($\lambda_{Edd} \gtrsim$ 0.001) they become 1-5 Gyr. Using the estimations of $L_{BH}$ and $M_{BH}$ (see Figure \ref{fig:agn_mbh}) we calculate the $\lambda_{Edd}$ for our sample, finding that the AGN hosts with a metallicity decrease have a median $\lambda_{Edd}$ of 0.0017, putting them in the lower values of the Eddington ratio.

Our results show that the maximum metallicity occurred, on average, about 3.6 Gyr ago which is compatible with these estimations. This measurement does not necessarily imply that these AGN have been active for this period of time, galaxies with a period of activity which previously completely stopped and which has been re-activated recently also would fulfill the criteria.

Conversely, it is possible that many of the currently quiescent galaxies have experienced an AGN phase in the past which could have contributed to the presence of a metallicity decrease in their ChEH.

We also measure how the metallicity decreases change with the star forming status of the galaxies. We separate between star-forming galaxies (SFG), Green Valley galaxies (GVG) and retired galaxies (RG), defined using the value of EW$_{ \mathrm{H}\alpha }$ measured at the effective radius. Following \citet{Lacerda2020} we use 3\AA{} as the value below which galaxies are retired and 10\AA{} as the value above which galaxies are star-forming, with GVG in between.

Regarding the incidence of metallicity decreases in these sub-groups we find that RG and GVG show a higher incidence than SFG, with $p<1e-5$ using Student's t-test for the difference between RG and SFG.
Outflows due to massive stellar feedback are expected to occur in SFG \citep{Lilly2013} but it bears mention that the feature we are measuring is specifically a decrease in metallicity that is roughly maintained to the current value. In other words, we are biased towards galaxies that have not had significant enrichment after the metal loss. For RG and GVG to show this type of feature more than SFG is therefore expected to an extent.

In the process of detecting the presence of a decrease we also measured the gap in metallicity after the decrease for each such galaxy, defined as the difference between the maximum value of the metallicity and the currently observed one. In the same manner we measured the LBT interval between the time at which the galaxy had its maximum metallicity and the time at which the light we are currently observing was emitted. It is important to note that the absolute values for the metal loss are dependent on the stellar libraries employed in the analysis (see Camps-Farina et al 2021, in prep.). However, this does not affect the relative values of the decrease allowing us to compare them between groups of galaxies.

The average value of these properties for each subgroup is listed in Table \ref{tab:props}, calculated using both the mean and the median. Perhaps unexpectedly, we find no significant difference in the average values of these two parameters for quiescent and AGN host galaxies.

In contrast, there is a clear trend with the star-forming status of the galaxies by which the higher the level of star formation the higher the decrease in metallicity. In combination with the lower percentage of galaxies with a decrease for SFG we find that this group of galaxies has fewer galaxies with a decrease in metallicity but these are significantly deeper. There are two plausible explanations for this: inflows of pristine gas and stellar feedback driven winds. The popular "bath-tub" model \citep[][]{Lilly2013} explains the sustained star formation in galaxies via a steady stream of pristine gas from the halo, which would dilute the metals in the disc producing the drop in metallicity. The second explanation is that supernovae and massive star winds fueled by starbursts eject enriched material which overcomes the potential of the disc thus preventing its recycling \citep[e.g.,][]{Veilleux2005,Tanner2017,Lopez-Coba2019,Lopez-Coba2020}.

Additionally, it could be the effect of interactions or mergers between galaxies, which can funnel gas from the outskirts towards the center of galaxies and produce star formation. Since the gas at the outskirts tends to be less enriched than that of the center \citep[e.g.][]{GonzalezDelgado2014,Sanchez-Menguiano2018} the same effect of metal dilution would be observed.

 \begin{figure*}
 	\includegraphics[width=\linewidth]{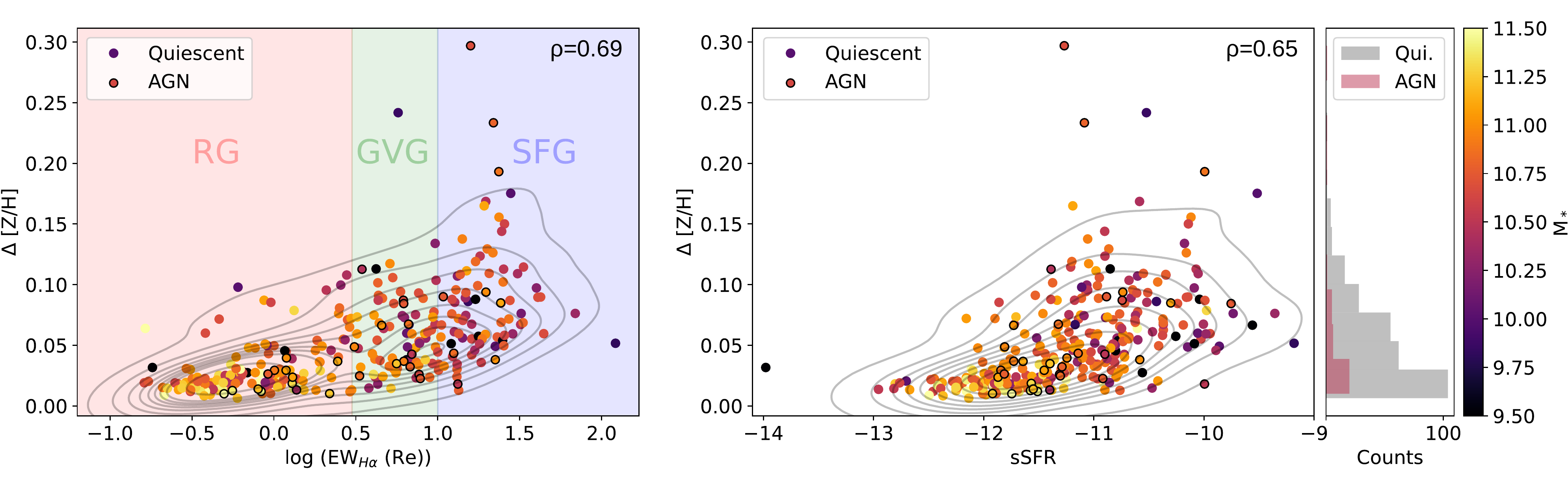}
     \caption{On the left, the relation between the equivalent width in H$\alpha$ and the decrease in metallicity, with the symbols differentiating between AGN hosts (black edges) and quiescent galaxies. The colored areas correspond to RG (red), GVG (green) and SFG (blue) defined by EW$_{H\alpha}$ (Re) cuts of 3 \AA{} and 10 \AA{}. On the right, the relation between the specific star formation rate (sSFR) and the decrease in metallicity, with the panel on the right showing a histogram of the distribution of values in the Y-axis. The colors of the symbols for both relations correspond to stellar mass. The upper right corner of each plot shows the Spearman correlation coefficient.}
     \label{fig:ew_dz}
 \end{figure*}

In Figure \ref{fig:ew_dz} we show the relation between the decrease in metallicity and both the EW$_{ \mathrm{H}\alpha }$ as well as the specific star formation rate (sSFR) obtained from the fossil record, for AGN hosts and quiescent galaxies. There is a clear trend between how star-forming a galaxy is (as currently observed) and the depth of the decrease in metallicity, with the galaxies with higher sSFR having a deeper drop. The presence of an AGN does not appear to have an effect on this relation, though the highest values of the decrease appear mostly on AGN.

The relation between sSFR and the amount of metals lost is evidence for the presence of outflows related to star-formation which supports that the metallicity decrease is driven by supernovae and stellar winds. These should dominate the SFG sub-group and indeed the relation steepens at the high sSFR range compared to the low sSFR one which corresponds to RG. It is important to note that this relation is not representative of the SFG sample in general, only of the portion that shows a decrease in metallicity. The reason it is important is that star-formation promotes both metal production and outflows so a positive trend between SFR and metal loss could be interpreted such that outflows outpace enrichment in general. The analysis implicitly selects galaxies where the metal loss overcomes the enrichment, as galaxies where the opposite happens would not show the metal decrease and therefore they are not selected.

The evidence for star-formation driven outflows also affects the interpretation of $\Delta$LBT for AGN hosts. $\Delta$LBT is not a direct measurement of when the AGN would have been "turned on" but simply how long ago the galaxy had its maximum metallicity. The higher incidence of metallicity decreases in the AGN host sample implies star-formation driven outflows are not enough to explain the results for the AGN sample, but we can expect previous star-formation driven outflows to bias the average $\Delta$LBT in AGN hosts towards higher values.

\acknowledgments

We are grateful for the support of a CONACYT grant CB-285080 and FC-2016-01-1916, and funding from the PAPIIT-DGAPA-IN100519, PAPIIT-DGAPA-IN103820 and PAPIIT-DGAPA-IA100420 (UNAM) projects.
J.B-B acknowledges support funding from the CONACYT grant CF19-39578.
 L.G. acknowledges financial support from the Spanish Ministry of Science, Innovation and Universities (MICIU) under the 2019 Ram\'on y Cajal program RYC2019-027683 and from the Spanish MICIU project PID2020-115253GA-I00.
We are grateful to the referee for very helpful comments on the manuscript.
 
\vspace{5mm}


\software{astropy \citep{AstropyCollaboration2013, AstropyCOllaboration2018},  
          }




\bibliography{agn_califa}{}
\bibliographystyle{aasjournal}



\end{document}